\documentclass[preprint,pteplogo]{ptephy_v2}

\preprintnumber{YITP-25-161, J-PARC-TH-0326} 
\usepackage{amsmath,amssymb}
\usepackage[colorlinks=true,allcolors=blue,breaklinks]{hyperref}
\usepackage{comment}
\usepackage{color}

\newcommand{\llangle}{\mathopen{\langle\!\langle}}
\newcommand{\rrangle}{\mathclose{\rangle\!\rangle}}

\newcommand{\Bigllangle}{\mathopen{\Big\langle\!\!\Big\langle}}

\newcommand{\Bigrrangle}{\mathclose{\Big\rangle\!\!\Big\rangle}}

\newcommand{\sumint}[1]{%
  \mathchoice%
    {\displaystyle\sum\!\!\!\!\!\!\!\!\int\Big._{#1\,}}
    {\textstyle\sum\!\!\!\!\!\!\raisebox{-0.2ex}{\scalebox{1.3}{$\int$}}\big._{#1}}
    {\scriptstyle\sum\!\!\!\int\!_{#1}}
    {\scriptscriptstyle\sum\!\!\!\int\!_{#1}}
}



\begin{document}

\title{Efficiency correction of particle-averaged quantities}


\author[1,2]{Masakiyo Kitazawa}
\author[3]{ShinIchi Esumi}
\author[3]{Takafumi Niida}
\author[3]{Toshihiro Nonaka}

\affil[1]{Yukawa Institute for Theoretical Physics, Kyoto University, Kyoto, 606-8502, Japan \email{kitazawa@yukawa.kyoto-u.ac.jp}}
\affil[2]{J-PARC Branch, KEK Theory Center, Institute of Particle and Nuclear Studies, KEK, Tokai, Ibaraki, 319-1106, Japan}

\affil[3]{Tomonaga Center for the History of the Universe, University of Tsukuba, Tsukuba, Ibaraki, 305-0006, Japan}

\begin{abstract}%
We derive analytic formulas to reconstruct particle-averaged quantities from experimental results that suffer from the efficiency loss of particle measurements. These formulas are derived under the assumption that the probabilities of observing individual particles are independent. The formulas do not agree with the conventionally used intuitive formulas. 
\end{abstract}

\subjectindex{xxxx, xxx}

\maketitle

\section{Introduction}
\label{sec:intro}

Consider a collider experiment that produces many ``collision events,'' where in each event many particles are produced. Every particle carries physical quantities, such as mass, momentum, and various charges. Suppose that we are interested in their \textit{average over particles} of a specific species \textit{in each collision event}. More specifically, denoting the particle number in a collision event by $N$ and the physical quantity carried by the $i$th particle by $\xi_i$ ($i=1,\cdots,N)$, the quantity of interest is 
\begin{align}
    \frac1N \sum_{i=1}^N \xi_i ,
    \label{eq:aver}
\end{align}
while the \textit{sum} of the physical quantity is given by
\begin{align}
    \sum_{i=1}^N \xi_i .
    \label{eq:sum}
\end{align}
Our goal is to obtain the average of Eq.~\eqref{eq:aver} \textit{over collision events} and its event-by-event distribution.

Such measurements are quite common in relativistic heavy-ion collisions~\cite{Yagi:2005yb,Braun-Munzinger:2015hba,Foka:2016vta,Song:2017wtw,Shen:2020mgh,Chen:2024aom}. For example, the mean transverse momentum $p_T$ and the flow-anisotropy parameters such as the directed and elliptic flows~\cite{Voloshin:2008dg} belong to this category. In addition to event-by-event averages, their event-by-event fluctuations have recently received much attention~\cite{ATLAS:2013xzf,CMS:2017glf,ALICE:2018rtz}. For instance, higher-order correlation functions of the flow-anisotropy parameters $v_n$ and the mean $p_T$ have been actively investigated as observables sensitive to the shape of colliding nuclei~\cite{Bozek:2016yoj,giacalone2020,Jia:2021qyu,bally2022,Jia:2022ozr,Hagino:2025vxe,ATLAS:2022dov,STAR:2024wgy}.

In real experiments, the performance of experimental detectors is always imperfect. For example, every particle arriving at a detector is observed only with a probability less than unity, called the efficiency. Modifications of the observed results due to these detectors' effects must be corrected to recover the true results. In the present study, we examine this procedure for quantities of the form Eq.~\eqref{eq:aver}, focusing on the efficiency effects, i.e., loss of particles in their measurements. Throughout this paper, we call this procedure the efficiency correction.

Efficiency corrections of Eq.~\eqref{eq:sum}
have been well established. The efficiency-corrected result of Eq.~\eqref{eq:sum} is simply expressed in terms of the observed results with the efficiency loss as $\sum_i \xi_i/r_i$, where $r_i$ denotes the efficiency of the $i$th particle and the sum runs over the observed particles. While the efficiency correction of the higher-order correlations of Eq.~\eqref{eq:sum} is more involved, the analytic treatment for their correction has been developed in the literature~\cite{Kitazawa:2011wh,Kitazawa:2012at,Bzdak:2012ab,Luo:2014rea,Asakawa:2015ybt,Kitazawa:2016awu,Bzdak:2016qdc,Nonaka:2016xje,Nonaka:2017kko,Kitazawa:2017ljq}. See Ref.~\cite{Nonaka:2018mgw} for the general argument. On the other hand, analytic procedures for the efficiency correction of the {\it particle-averaged} quantities, Eq.~\eqref{eq:aver}, have not been addressed in the literature to the best of the authors' knowledge. Although there exist formulas that are conventionally used in experimental analyses (Eq.~\eqref{eq:conventional}), see, for example, Refs.~\cite{ATLAS:2013xzf,ATLAS:2022dov,STAR:2024wgy}, the authors are not aware of any literature in which their derivation is given. Unfolding methods~\cite{Jia:2013tja,Esumi:2020xdo} are alternative ways to deal with the correction. However, they typically require large numerical resources, and their reconstruction procedures are generally less transparent.

In the present study, we derive analytic formulas for the efficiency correction of quantities of the form Eq.~\eqref{eq:aver}, as well as their higher-order correlations. 
These formulas are derived under the assumption that the detection probabilities of individual particles are mutually uncorrelated. This assumption provides a straightforward yet effective framework for incorporating efficiency loss. For example, the efficiency correction of the net-proton number cumulants reported recently in Ref.~\cite{STAR:2025zdq} was carried out with this assumption. The formulas thus will be applicable to various experimental analyses, whereas residual corrections may be required due to the violation of this assumption, which should be carefully evaluated for each experimental group~\cite{STAR:2021iop,HADES:2020wpc,ALICE:2022xpf}. We also show that these results do not agree with the conventional formulas Eq.~\eqref{eq:conventional}.

This paper is organized as follows. In Sec.~\ref{sec:setting}, we clarify the problem that is addressed in this study and summarize the main results. In Sec.~\ref{sec:1st}, we address the efficiency correction of the particle-averaged quantities of the form Eq.~\eqref{eq:aver}. We then extend the analysis to higher-order correlations in Sec.~\ref{sec:higher}. The last section is devoted to discussions of these results. In App.~\ref{app:1/N}, we discuss the analytic procedure of the efficiency correction in a simple model. In App.~\ref{app:binom}, we demonstrate the validity of our formulas in a simple model.

\section{Definition of the problem and summary of the results}
\label{sec:setting}

\subsection{Quantities to be investigated}

In this study, besides the quantities of the form Eq.~\eqref{eq:aver}, we derive analytic formulas for the efficiency correction of quantities 
\begin{align}
    &\frac1{N(N-1)} \sum_{i\ne j} \xi^{(1)}_i \xi^{(2)}_j , 
    \label{eq:corr2}
    \\
    &\frac1{N(N-1)(N-2)} \sum_{i\ne j,j\ne k,k\ne i} \xi^{(1)}_i \xi^{(2)}_j \xi^{(3)}_k, 
    \label{eq:corr3}
\end{align}
etc., where $\xi^{(w)}_i$ represent physical quantities, with the superscript $w$ indicating different types of quantities, although they may be identical. We refer to Eqs.~\eqref{eq:corr2} and~\eqref{eq:corr3} as the second- and third-order correlations, respectively. Here, the sum over the subscripts, $i,j,k$, is taken over particles produced in a collision event, whose number is $N$. In Eqs.~\eqref{eq:corr2} and~\eqref{eq:corr3}, the self-correlations, i.e., the contribution of the identical particle, $i=j$, are excluded from the summation, because they are usually excluded in the analyses of flow harmonics in relativistic heavy-ion collisions. For example, the conventional definition of the flow-anisotropy parameters from the two-particle correlations is~\cite{Voloshin:1994mz,Borghini:2000sa,STAR:2002hbo,PHENIX:2002hqx}
\begin{align}
    v_n^2\{2\} = \frac1{N(N-1)} \sum_{i\ne j} e^{in(\theta_i-\theta_j)} ,
\end{align}
where $\theta_i$ is the azimuthal angle of the $i$th particle. 
The formulas of the efficiency correction for higher-order correlations including the self-correlations, for example $(\sum_i \xi_i /N)^2$,
will be presented in a future publication. 

\subsection{Assumptions and results}

In this study, we derive analytic formulas for the efficiency correction of quantities of the form Eqs.~\eqref{eq:aver},  \eqref{eq:corr2}, 
and~\eqref{eq:corr3}. 
Major assumptions imposed in this analysis are as follows:
\begin{itemize}
    \item 
    The distribution of the particle number in each collision event and the values of physical quantities carried by them are expressed by a classical probability distribution function. For the simplest case, the distribution function is represented by $P(N;\vec\xi\,)$, where $N$ is the number of produced particles in each event and $\vec\xi=(\xi_1,\cdots,\xi_N)$ are the values of a physical quantity carried by individual particles. No constraints on the form of $P(N;\vec\xi\,)$ are imposed.
    \item 
    The detector observes every produced particle with some probability, i.e., efficiency. Moreover, the probabilities of observing individual particles are mutually uncorrelated. 
    \item 
    The values of the efficiencies can differ from particle to particle, but they are specified for all observed particles. We denote the efficiency of the $i$th particle as $r_i$ unless otherwise stated.
    \item 
    We do not consider other detectors' effects. For example, the values of the physical quantities $\xi_i$ may be measured with experimental errors. There will also be misidentification of non-existing particles. These effects, however, are not considered in this study for simplicity.
\end{itemize}

Our final results, i.e., analytic formulas for the efficiency corrections of Eqs.~\eqref{eq:aver}, \eqref{eq:corr2}, and~\eqref{eq:corr3} obtained under these assumptions, are given in Eqs.~\eqref{eq:correction3}, \eqref{eq:correctionC2}, and~\eqref{eq:correctionC3}.

In the literature, the following conventional formulas are sometimes used for the efficiency correction~\cite{ATLAS:2013xzf,ATLAS:2022dov,STAR:2024wgy}
\begin{align}
    &\Big\langle \frac{\sum_i\xi_i}N \Big\rangle 
    = \Bigllangle \frac{\sum_i\xi_i/r_i}{\sum_i 1/r_i} \Bigrrangle,
    \qquad
    \Big\langle \frac{\sum_{i\ne j} \xi_i^{(1)} \xi_j^{(2)}}{N(N-1)} \Big\rangle 
    = \Bigllangle \frac{\sum_{i\ne j} \xi_i^{(1)} \xi_j^{(2)}/(r_ir_j)}{\sum_{i\ne j} 1/(r_ir_j)} \Bigrrangle,
    \notag \\
    &\Big\langle \frac{\sum_{i\ne j,j\ne k,k\ne i} \xi_i^{(1)} \xi_j^{(2)} \xi_k^{(3)}}{N(N-1)(N-2)} \Big\rangle 
    = \Bigllangle \frac{\sum_{i\ne j,j\ne k,k\ne i} \xi_i^{(1)} \xi_j^{(2)} \xi_k^{(3)}/(r_ir_jr_k)}{\sum_{i\ne j,j\ne k,k\ne i} 1/(r_ir_jr_k)} \Bigrrangle,
    \label{eq:conventional}
\end{align}
etc., where $\langle\cdot\rangle$ denotes the true expectation value over the collision events without efficiency loss, while $\llangle\cdot\rrangle$ means the expectation value taken over experimentally observed particles after the efficiency loss. However, to the best of the authors' knowledge no literature has derived these formulas. The results obtained in the present study, Eqs.~\eqref{eq:correction3}, \eqref{eq:correctionC2}, and~\eqref{eq:correctionC3}, do not agree with Eq.~\eqref{eq:conventional} even at the first order for Eq.~\eqref{eq:aver}.

\section{Correction for averages}
\label{sec:1st}

In this section, we address the efficiency correction of the event-by-event average of quantities of the form Eq.~\eqref{eq:aver}. We recommend that the readers consult App.~\ref{app:1/N} before this section, which will help in understanding the manipulations in this section.

\subsection{Uniform efficiency}
\label{sec:1st_uniform}

We begin with the simple case where the detector's efficiency is uniform and all particles are observed with a single common efficiency $r$ ($0<r<1$).

In this case, the experimental result in each collision event is specified by the number of particles $N$ and the physical quantities carried by individual particles $\xi_i$ ($i=1,\cdots,N$). We denote their probability distribution as 
\begin{align}
    P(N;\vec{\xi}\,) 
    = P_{\rm N}(N) p_N(\vec\xi\,) ,
    \label{eq:P1}
\end{align}
which satisfies the normalization conditions $\sum_N \int d^N\vec\xi P(N;\vec\xi\,)=\sum_N P_{\rm N}(N) = \int d^N\vec\xi\, p_N(\vec\xi\,)=1$ with $\vec\xi=(\xi_1,\cdots,\xi_N)$. It is reasonable to require that Eq.~\eqref{eq:P1} is invariant under the permutation of particles, e.g., $p_N(\xi_1,\xi_2,\cdots,\xi_N)=p_N(\xi_2,\xi_1,\cdots,\xi_N)$, etc. We also impose the condition $P_{\rm N}(0)=0$ so that the division by $N$ in Eq.~\eqref{eq:aver} is always well defined. We denote the sum of the physical quantity as 
\begin{align}
    Q = \sum_{i=1}^N \xi_i .
\end{align}
The true expectation value of the average, Eq.~\eqref{eq:aver}, then reads
\begin{align}
    \Big\langle \frac QN \Big\rangle 
    = \sumint{P} \frac QN ,
    &&
    \sumint{P} \equiv \sum_N \int d^N \vec\xi\, P(N;\vec\xi\,) ,
    \label{eq:<Q/N>sumint}
\end{align}
where $\langle \cdot \rangle$ denotes the average over the collision events as defined in Eq.~\eqref{eq:<Q/N>sumint}.

We now consider the experimental measurement of this system with the efficiency loss. We denote the number of observed particles as $n$, the sum of the physical quantities over the observed particles as
\begin{align}
    q = \sum_{i\in({\rm observed})} \xi_i ,
\end{align}
and the probability distribution function to find $n$ and $q$ in a collision event as $\tilde P(n,q)$.

To relate $\tilde P(n,q)$ with $P(N;\vec\xi\,)$, it is convenient to introduce a set of variables representing the success of observing individual particles
\begin{align}
    b_i 
    = \begin{cases}
        1 & \textrm{(observed)} \\
        0 & \textrm{(not observed)}
    \end{cases}
    \quad (i=1,\cdots, N).
    \label{eq:b_i}
\end{align}
Using Eq.~\eqref{eq:b_i}, one can write $n = \sum_{i=1}^N b_i$ and $q=\sum_{i=1}^N b_i \xi_i$. Since the probabilities of having $b_i=0$ and $1$ are $1-r$ and $r$, respectively, one can express $\tilde P(n,q)$ using $P(N;\vec\xi\,)$ as 
\begin{align}
    \tilde P (n,q) 
    &= \sumint P \prod_{i=1}^N \sum_{b_i=0}^1 (1-r)^{1-b_i}r^{b_i} \cdot \delta_{n,\sum_i b_i} \delta\Big( q - \sum_i b_i \xi_i \Big)
    \notag \\
    &= \sumint{P} \sum_{\{b_i\}} \Big[\prod_{i=1}^N(1-r)^{1-b_i}r^{b_i}\Big] \delta_{n,\sum_i b_i} \delta\Big( q - \sum_i b_i \xi_i \Big)
    \notag \\
    &= \sumint{P} \sum_{\{b_i\}} (1-r)^{N-n} r^n \delta_{n,\sum_i b_i} \delta\Big( q - \sum_i b_i \xi_i \Big) ,
    \label{eq:P1tilde}
\end{align}
where $\sum_{\{b_i\}}$ means the summation over all possible combinations of $b_i=(0,1)$. 

To proceed further, we introduce the generating function
\begin{align}
    \tilde G(s,\theta) 
    = \sum_n \int dq \tilde P(n,q) s^n e^{q\theta} 
    \equiv \sumint{\tilde P} s^n e^{q\theta} 
    \equiv \llangle s^n e^{q\theta} \rrangle ,
    \label{eq:G1tilde}
\end{align}
where the double braket $\llangle \cdot \rrangle$ denotes the expectation value for observed particles as defined in Eq.~\eqref{eq:G1tilde}.
Substituting Eq.~\eqref{eq:P1tilde} into Eq.~\eqref{eq:G1tilde}, we obtain
\begin{align}
    \tilde G(s,\theta) 
    &= \sumint{P} \sum_{\{b_i\}} \Big[\prod_i (1-r)^{1-b_i}r^{b_i}\Big] s^{\sum_i b_i} e^{\theta\sum_i b_i \xi_i}
    \notag \\
    &= \sumint{P} \sum_{\{b_i\}} \prod_i (1-r)^{1-b_i}(rse^{\xi_i\theta})^{b_i} 
    \notag \\
    &= \sumint{P} \prod_i ( 1-r+rse^{\xi_i\theta} ) .
    \label{eq:G1tilde2}
\end{align}
The $\theta$ derivative of Eq.~\eqref{eq:G1tilde2} is given by
\begin{align}
    \frac{\partial \tilde G(s,\theta)}{\partial\theta}
    &= \sumint{P} rs \sum_i \xi_i e^{\xi_i\theta} \prod_{j\ne i} ( 1-r+rs e^{\xi_j\theta} ) ,
    \label{eq:dG1a} \\
    \frac{\partial \tilde G(s,\theta)}{\partial\theta}\Big|_{\theta=0}
    &= \sumint{P} Q rs ( 1-r+rs)^{N-1} 
    = \sumint{P} Q r^N s (s-\alpha)^{N-1} ,
    \label{eq:dG1b}
\end{align}
with $\alpha\equiv (r-1)/r$.

From Eq.~\eqref{eq:dG1b}, we find
\begin{align}
    \int_\alpha^1 ds \frac1s \frac{\partial \tilde G(s,\theta)}{\partial\theta}\Big|_{\theta=0}
    = \sumint{P} Q r^N \int_\alpha^1 ds (s-\alpha)^{N-1} 
    = \sumint{P} \frac QN = \Big\langle \frac QN \Big\rangle.
    \label{eq:<Q/N>}
\end{align}
On the other hand, a similar manipulation on both sides of Eq.~\eqref{eq:G1tilde} leads to
\begin{align}
    \frac{\partial \tilde G(s,\theta)}{\partial\theta} \Big|_{\theta=0}
    &= \sumint{\tilde P,n\ne0} q s^n ,
    \label{eq:dG1c} \\
    \int_\alpha^1 ds \frac1s \frac{\partial \tilde G(s,\theta)}{\partial\theta} \Big|_{\theta=0}
    &= \sumint{\tilde P,n\ne0} q \int_\alpha^1 ds s^{n-1} 
    = \sumint{\tilde P,n\ne0} \frac qn ( 1- \alpha^n )
    = \Bigllangle \frac qn (1-\alpha^n) \Bigrrangle_{n\ne0} .
    \label{eq:<q/n>}
\end{align}
On the right-hand side of Eq.~\eqref{eq:dG1c}, we removed the contribution of $n=0$ to the summation because it trivially vanishes as $q=0$ in this case. 

Comparing Eqs.~\eqref{eq:<Q/N>} and~\eqref{eq:<q/n>}, one arrives at 
\begin{align}
    \Big\langle \frac QN \Big\rangle = \Bigllangle \frac qn (1-\alpha^n) \Bigrrangle_{n\ne0}.
    \label{eq:correction1}
\end{align}
In Eq.~\eqref{eq:correction1}, $\langle Q/N \rangle$ of the true distribution $P(N;\vec\xi\,)$ is represented by the observed quantities. Therefore, Eq.~\eqref{eq:correction1} is the formula to reconstruct the true value of $\langle Q/N \rangle$ from the experimental results with efficiency loss. Here, the expectation value $\llangle \cdot \rrangle_{n\ne0}$ indicates that zero is substituted for contributions from events with $n=0$, as defined in Eq.~\eqref{eq:<q/n>}. We emphasized that this does not mean that the $n=0$ events are excluded from the event ensemble.

It is notable that the right-hand side of Eq.~\eqref{eq:correction1} is not $\llangle q/n\rrangle$, but has an additional term $-\llangle (q/n)\alpha^n \rrangle$. The appearance of this term is attributed to the fact that the $n=0$ contribution is removed on the right-hand side of Eq.~\eqref{eq:correction1}. To show this, we rewrite this term as 
\begin{align}
    -\Bigllangle \frac qn \alpha^n \Bigrrangle_{n\ne0} 
    &= -\sum_{n\ne0} \int dq \frac qn \alpha^n \sumint{P} \sum_{\{b_i\}} (1-r)^{N-n} r^n \delta_{n,\sum_i b_i} \delta(q-\sum_i b_i \xi_i)
    \notag \\
    &= -\sumint{P} (1-r)^N \sum_{\{b_i\}} \sum_{n\ne0} \frac {\sum_i b_i\xi_i}n (-1)^n \delta_{n,\sum_i b_i} 
    \notag \\
    &= -\sum_N P_{\rm N}(N) (1-r)^N \langle \xi \rangle_N \sum_{n\ne0} {}_N C_n (-1)^n
    \notag \\
    &= \big\langle (1-r)^N \langle \xi \rangle_N \big\rangle.
    \label{eq:qnalpha}
\end{align}
Here, on the third line we defined the expectation value of $\xi_i$ for a given $N$ as $\langle \xi\rangle_N \equiv \int d^N \vec\xi p_N(\vec\xi\,) \sum_i \xi_i/N = \int d^N \vec\xi p_N(\vec\xi\,) \xi_i$. To obtain the last line, we used $\sum_{n\ne0} {}_N C_n (-1)^n=-1$. Using Eq.~\eqref{eq:qnalpha} and $\langle Q/N \rangle= \big\langle \langle \xi\rangle_N\big\rangle$, one can write
\begin{align}
    \Bigllangle \frac qn \Bigrrangle_{n\ne0}
    = \Big\langle \frac QN \Big\rangle - \big\langle (1-r)^N \langle \xi \rangle_N \big\rangle 
    = \big\langle \big(1-(1-r)^N\big) \langle \xi \rangle_N \big\rangle .
    \label{eq:<q/n>nne0}
\end{align}
Since $(1-r)^N$ is the probability of having $n=0$ in events with $N$ produced particles, the above calculation shows that $\llangle (q/n)\alpha^n \rrangle$ represents the contribution of these events.

We notice that $\alpha$ is negative, and $|\alpha|\ge1$ for $r\le1/2$. Therefore, $|\alpha^n|$ is divergent for $n\to\infty$ with flipping signs for $r<1/2$, which in practice would result in an unstable result of the sum on the right-hand side of Eq.~\eqref{eq:correction1} for $r<1/2$. This may mean that a proper analysis of $\langle Q/N \rangle$ is difficult for $r<1/2$. On the other hand, the term $\llangle (q/n)\alpha^n \rrangle$ may be negligible when the probability of observing no particles is well suppressed even for $r<1/2$. It is not clear when this term causes large errors in the reconstruction.

Another notice concerning this term is that $\llangle \alpha^n \rrangle=0$ is generally satisfied from the definition of $\tilde P(n,q)$. This property may be used to modify Eq.~\eqref{eq:correction1} to improve its convergence.

\subsection{Multiple efficiencies}
\label{sec:1st_mult}

Next, we consider a more general problem where the efficiency of our detector is not uniform. 

We start from the case that our detector has distinct regions having different efficiencies, $r_1,\cdots,r_{A}$, which we call the efficiency bins in what follows, with $A$ being the number of bins. 
In this case, the number of particles entering each efficiency bin must be specified to characterize the collision events. We thus introduce the true probability distribution function 
\begin{align}
    P(\{N;\vec\xi\,\}) = P(N_1,\cdots,N_A;\vec\xi^{\,(1)},\cdots,\vec\xi^{\,(A)}) ,
\end{align}
where $N_a$ ($a=1,\cdots,A$) is the number of particles that enter the efficiency bin $a$, and $\vec\xi^{\,(a)}=(\xi^{(a)}_1,\cdots,\xi^{(a)}_{N_a})$ represents the physical quantities carried by these particles. The total number of particles, $N$, and the physical quantity, $Q$, in each event are given by
\begin{align}
    N = \sum_{a=1}^A N_a, 
    &&
    Q = \sum_{a=1}^{A} Q^{(a)} = \sum_{a=1}^{A} \sum_{i=1}^{N_A} \xi_i^{(a)} ,
\end{align}
with $Q^{(a)} \equiv \sum_{i=1}^{N_A} \xi_i^{(a)}$.

As a result of the measurement of this system with the efficiency loss, we may observe $n_1,\cdots,n_A$ particles and the sum of physical quantities $q^{(1)},\cdots,q^{(A)}$ in individual efficiency bins, respectively, with $q^{(a)}=\sum_{i\in\textrm{(observed)}}\xi_i^{(a)}$. We denote their probability distribution function as $\tilde P(\{n,q\})=\tilde P(n_1,\cdots,n_A;q^{(1)},\cdots,q^{(A)})$. Repeating the same procedure as the previous subsection, one can relate $\tilde P(\{n,q\})$ with $P(\{N;\vec\xi\,\})$ as follows:
\begin{align}
    \tilde P(\{n,q\})
    &= \sumint{P} \sum_{\{b_i^{(a)}\}} \prod_{i=1}^{N_a} (1-r_a)^{1-b_i^{(a)}} r_a^{b_i^{(a)}} \delta_{n_a,\sum_ib_i^{(a)}} \delta\Big( q^{(a)} - \sum_i b_i^{(a)} \xi_i^{(a)} \Big) ,
\end{align}
where $\sumint{P}$ now means 
\begin{align}
    \sumint{P} &= \sum_{N_1,\cdots,N_A} \int \prod_{a=1}^A d^{N_a} \vec\xi^{\,(a)} P(\{N;\vec\xi\,\}) ,
\end{align}
and the meaning of $\sum_{\{b_i^{(a)}\}}$ is understood.

To proceed with the manipulation further, we introduce the generating function 
\begin{align}
    \tilde G(\{ s_a,\theta_a\} )
    &= \tilde G(s_1,\cdots,s_A,\theta_1,\cdots,\theta_A)
    \notag \\
    &= \sumint{\tilde P} \prod_{a=1}^A s_a^{n_a} e^{\theta_a q_a}
    \label{eq:G2a} \\
    &= \sumint{P} \prod_{a=1}^A \prod_{i=1}^{N_a} (1-r_a+r_a s_a e^{\theta_a\xi_i^{(a)}}) ,
    \label{eq:G2b}
\end{align}
whose arguments are now increased in response to $\{n_a\}$ and $\{q^{(a)}\}$.
As the $\theta_a$ derivative of Eq.~\eqref{eq:G2b} reads
\begin{align}
    \partial_a \tilde G
    &\equiv \frac{\partial\tilde G}{\partial \theta_a} 
    \notag \\
    &= \sumint{P} r_a s_a \sum_i \xi_i^{(a)} e^{\theta_a \xi_i^{(a)}} \prod_{j\ne i} (1-r_a + r_a s_a e^{\theta_a \xi_j^{(a)}} ) 
    \prod_{b\ne a} \prod_k (1-r_b+r_b s_b e^{\theta_b\xi_k^{(b)}}),
\end{align}
one obtains
\begin{align}
    \frac{\partial_a \tilde G}{s_a} \Big|_{\theta=0}
    &= \sumint{P} \frac{Q^{(a)} }{s_a-\alpha_a} \prod_{b=1}^A r_b^{N_b} (s_b-\alpha_b)^{N_b},
    \label{eq:dG2}
\end{align}
with $\alpha_a=(r_a-1)/r_a$.
We then introduce a new variable $\sigma$, which is related to $s_a$'s as 
\begin{align}
    s_a &= (1-\alpha_a) \sigma + \alpha_a = \frac\sigma{r_a} + \alpha_a ,
    \label{eq:s-sigma}
    \\
    \sigma &= \frac{s_a-\alpha_a}{1-\alpha_a} = r_a ( s_a -\alpha_a) .
    \label{eq:sigma-s}
\end{align}
and regard Eq.~\eqref{eq:dG2} as a function of the single variable $\sigma$ as
\begin{align}
    \frac{\partial_a \tilde G}{s_a} \Big|_{\theta=0}
    = \frac{\partial_a \tilde G\big(\frac\sigma{r_1}+\alpha_1,\cdots,\frac\sigma{r_A}+\alpha_A;0,\cdots,0\big)}{\sigma/r_a+\alpha_a} 
    &= \sumint{P} Q^{(a)} r_a \sigma^{\sum_a N_a -1} ,
\end{align}
which yields
\begin{align}
    \sum_{a=1}^A \frac{\partial_a \tilde G}{r_a s_a} \Big|_{\theta=0}
    &= \sumint{P} Q \sigma^{N-1}.
    \label{eq:sum_dG}
\end{align}

By taking the integral of both sides of Eq.~\eqref{eq:sum_dG}, one obtains
\begin{align}
    \int_0^1 d\sigma \sum_{a=1}^A \frac{\partial_a \tilde G}{r_a s_a} \Big|_{\theta=0}
    &= \sumint{P} \frac QN = \Big\langle \frac QN \Big\rangle.
    \label{eq:<Q/N>2}
\end{align}
Applying the same manipulation to Eq.~\eqref{eq:G2a}, we obtain
\begin{align}
    \partial_a \tilde G |_{\theta=0}
    &= \sumint{\tilde P,n\ne0} q^{(a)} \prod_{b=1}^A s_b^{n_b} ,
    \\
    \sum_{a=1}^A \frac{\partial_a \tilde G}{r_a s_a} \Big|_{\theta=0}
    &= \sumint{\tilde P,n\ne0} \sum_{a=1}^A \frac{q^{(a)}}{r_a (\sigma/r_a+\alpha_a)} \prod_{b=1}^A \Big( \frac\sigma{r_b} + \alpha_b \Big)^{n_b} ,
    \\
    \int_0^1 d\sigma \sum_{a=1}^A \frac{\partial_a \tilde G}{r_a s_a} \Big|_{\theta=0}
    &= \sumint{\tilde P,n\ne0} \sum_{a=1}^A \frac{q^{(a)}}{r_a} \int_0^1 d\sigma \frac1{\sigma/r_a+\alpha_a} \prod_{b=1}^A \Big( \frac\sigma{r_b} + \alpha_b \Big)^{n_b} ,
    \notag \\
    &= \Bigllangle \sum_{a=1}^A \frac{q^{(a)}}{r_a} \int_0^1 d\sigma \frac1{\sigma/r_a+\alpha_a} \prod_{b=1}^A \Big( \frac\sigma{r_b} + \alpha_b \Big)^{n_b} \Bigrrangle_{n\ne0},
    \label{eq:intdG}
\end{align}
with $n=\sum_{a=1}^A n_a$.

Comparing Eqs.~\eqref{eq:<Q/N>2} and~\eqref{eq:intdG}, we obtain the correction formula for this case as 
\begin{align}
    \Big\langle \frac QN \Big\rangle
    &= \Bigllangle \sum_{a=1}^A q^{(a)} K_{1,a} \Bigrrangle_{n\ne0},
    \label{eq:correction2} \\
    K_{1,a} &= \frac1{r_a} \int_0^1 d\sigma \frac1{\sigma/r_a+\alpha_a} \prod_{b=1}^A \Big( \frac\sigma{r_b} + \alpha_b \Big)^{n_b} 
    \notag \\
    &= \frac1{r_a} \int_0^1 d\sigma \Big( \frac\sigma{r_a} + \alpha_a \Big)^{n_a-1}  \prod_{b\ne a} \Big( \frac\sigma{r_b} + \alpha_b \Big)^{n_b} ,
\end{align}
where $K_{1,a}$ are dependent on $\{n_a\}$ and thus is inside the expectation value in Eq.~\eqref{eq:correction2}.

In real experiments, the values of efficiencies may be different particle by particle. In this case, the efficiency bins in Eq.~\eqref{eq:correction2} are divided into individual observed particles and 
one can rewrite Eq.~\eqref{eq:correction2} as
\begin{align}
    \Big\langle \frac QN \Big\rangle
    &= \Bigllangle \sum_{i=1}^n \xi_i k_{1;i} \Bigrrangle_{n\ne0} ,
    \label{eq:correction3}
    \\
    k_{1;i} &= \frac1{r_i} \int_0^1 d\sigma \prod_{j\ne i} \frac{\sigma + r_j\alpha_j}{r_j} = \Big( \prod_i r_i \Big)^{-1} \int_0^1 d\sigma \prod_{j\ne i}( \sigma + r_j -1 ),
    \label{eq:k1}
\end{align}
where $i$ runs over all observed particles and $r_i$ denotes the efficiency of the $i$th observed particle with $\alpha_i=(r_i-1)/r_i$. This formula would be more convenient in practical analyses.

\subsection{Comments}
\label{sec:1st_comment}

Several comments on Eqs.~\eqref{eq:correction2} and~\eqref{eq:correction3} are in order.

First, Eq.~\eqref{eq:correction3} does not agree with the conventional formula, the first equation in Eq.~\eqref{eq:conventional}. At the moment, the authors do not fully understand their mutual relation. For example, it is expected that they agree with each other in some limit, such as $N\to\infty$ or $r_i\to1$. However, the authors have not succeeded in finding such relations. 

Second, in App.~\ref{app:binom} we apply Eq.~\eqref{eq:correction3} to a simple model and show that the true value is reconstructed correctly, while the first equality of Eq.~\eqref{eq:conventional} cannot.

The final comment is concerned with the calculation of $k_{1;i}$ in Eq.~\eqref{eq:k1}. The simplest way to calculate $k_{1;i}$ is to evaluate the integral numerically. However, such a treatment may be numerically demanding especially when the particle and event numbers are huge. To reduce the numerical costs, one may use the relation 
\begin{align}
    k_{1;i} 
    &= \Big( \prod_i r_i \Big)^{-1} \int_0^1 d\sigma \sum_{m=0}^{n-1} e_m(\{r_j-1|j\ne i\}) \sigma^{n-m-1}
    \notag \\
    &= \Big( \prod_i r_i \Big)^{-1} \sum_{m=0}^{n-1} \frac{e_m(\{r_j-1|j\ne i\})}{n-m} ,
    \label{eq:k1S}
\end{align}
where $e_m(\{u_j\})\equiv \sum_{1\le j_1 < j_2 < \cdots <j_m\le n} u_{j_1}u_{j_2}\cdots u_{j_m}$ are the elementary symmetric polynomials of $\{u_i\}$. Then, the relations between the elementary symmetric polynomials, such as $e_m(\{u_j\})=e_m(\{u_j|j\ne i\})+ u_i e_{m-1}(\{u_j|j\ne i\})$, may be useful to calculate Eq.~\eqref{eq:k1S} for various $i$.

\section{Higher order correlations}
\label{sec:higher}

Next, we address the correction of the higher-order correlations, Eqs.~\eqref{eq:corr2} and~\eqref{eq:corr3}. We also extend our formalism to deal with multiple physical quantities.

\subsection{Uniform efficiency}
\label{sec:higher_uniform}

As a preliminary to this problem, we again begin with the simple case where the efficiency $r$ of the detector is common for all particles. We assume that individual particles carry multiple physical quantities $\xi_i^{(w)}$ ($w=1,2,\cdots$), where $w$ identifies different physical quantities. As before, $i$ labels different particles.
Notice that the meaning of the superscript of $\xi$ is different from that in Sec.~\ref{sec:1st_mult}, where it represents the efficiency bins. 

We denote the probability distribution function to find $N$ particles with physical quantities $\xi_i^{(w)}$ as 
\begin{align}
    P(N;\{\vec\xi^{\,(w)}\})
    = P(N;\vec\xi^{\,(1)},\vec\xi^{\,(2)},\cdots ).
\end{align}
The sum of the physical quantities in each event is given by
\begin{align}
    Q_w = \sum_{i=1}^N \xi_i^{(w)} .
\end{align}

After the measurement of this system with the efficiency $r$, we may observe $n$ particles and the sum of physical quantities $q_w=\sum_{i\in\textrm{(observed)}}\xi_i^{(w)}$. We denote the probability distribution function to find these results as $\tilde P(n;\{q_w\})=\tilde P(n;q_1,q_2,\cdots)$. By repeating the same procedure as Sec.~\ref{sec:1st_uniform}, we find that $\tilde P(n;\{q_w\})$ is related to $P(N;\{\vec\xi^{\,(w)}\})$ as 
\begin{align}
    \tilde P(n;\{q_w\}) 
    &= \sumint P \prod_{i=1}^N \sum_{b_i=0}^1 (1-r)^{1-b_i}r^{b_i} \cdot \delta_{n,\sum_i b_i} \sum_w \delta\Big( q_w - \sum_i b_i \xi_i^{(w)} \Big)
    \notag \\
    &= \sumint{P} \sum_{\{b_i\}} \Big[\prod_{i=1}^N(1-r)^{1-b_i}r^{b_i}\Big] \delta_{n,\sum_i b_i} \sum_w \delta\Big( q_w - \sum_i b_i \xi_i^{(w)} \Big) .
    \label{eq:P~m}
\end{align}

Next, following Sec.~\ref{sec:1st_uniform} we introduce the generating function
\begin{align}
    \tilde G(s,\{\theta_w\}) 
    = \tilde G(s,\theta_1,\theta_2,\cdots) 
    = \sumint{\tilde P} s^n \prod_w e^{q_w\theta_w} ,
    \label{eq:G~m}
\end{align}
where we introduced variables $\theta_1,\theta_2,\cdots$ in response to $q_1,q_2,\cdots$.
Plugging Eq.~\eqref{eq:P~m} into Eq.~\eqref{eq:G~m} yields
\begin{align}
    \tilde G(s,\{\theta_w\}) 
    &= \sumint{P} \sum_{\{b_i\}} \Big[\prod_i (1-r)^{1-b_i}r^{b_i}\Big] s^{\sum_i b_i} e^{\sum_w \theta_w \sum_i b_i \xi_i^{(w)}}
    \notag \\
    &= \sumint{P} \sum_{\{b_i\}} \prod_i (1-r)^{1-b_i} \big(rs e^{\theta_w\sum_w\xi_i^{(w)}}\big)^{b_i} 
    \notag \\
    &= \sumint{P} \prod_i \big( 1-r+rs e^{\sum_w \theta_w \xi_i^{(w)}} \big) .
    \label{eq:G~m2}
\end{align}
Then, the derivatives of Eq.~\eqref{eq:G~m2} are given by
\begin{align}
    \frac{\partial \tilde G(s,\{\theta\})}{\partial\theta_1}
    &= \sumint{P} rs \sum_i \xi_i^{(1)} e^{\sum_w \theta_w \xi_i^{(w)}} \prod_{j\ne i}\big( 1-r+rs e^{\sum_w \theta_w \xi_j^{(w)}} \big) ,
    \label{eq:d1G~m} \\
    \frac{\partial \tilde G(s,\{\theta\})}{\partial\theta_1\partial\theta_2}
    &= \sumint{P} rs \sum_i \xi_i^{(1)} \xi_i^{(2)} e^{\sum_w \theta_w \xi_i^{(w)}} \prod_{j\ne i}\big( 1-r+rs e^{\sum_w \theta_w \xi_j^{(w)}} \big) 
    \notag \\
    &\phantom{=} + \sumint{P} r^2s^2 \sum_{i\ne j} \xi_i^{(1)} \xi_j^{(2)} e^{\sum_w \theta_w \xi_i^{(w)}} \prod_{k\ne i,k\ne j}\big( 1-r+rs e^{\sum_w \theta_w \xi_k^{(w)}} \big) ,
    \label{eq:d2G~m}
\end{align}
and so forth. Substituting $\theta=0$ to these results, one obtains
\begin{align}
    \frac{\partial \tilde G}{\partial\theta_1}\Big|_{\theta=0}
    &= \sumint{P} Q_1 r^N s( s-\alpha )^{N-1} ,
    \label{eq:d1G~m0} \\
    \frac{\partial^2 \tilde G}{\partial\theta_1 \partial\theta_2}\Big|_{\theta=0}
    &= \sumint{P} r^N \Big[ Q_{12} s ( s-\alpha )^{N-1} 
    + \{ Q_1 Q_2 \} s^2 ( s-\alpha )^{N-2} \Big] ,
    \label{eq:d2G~m0} \\
    \frac{\partial^3 \tilde G}{\partial\theta_1 \partial\theta_2 \partial\theta_3}\Big|_{\theta=0}
    &= \sumint{P} r^N \Big[ Q_{123} s ( s-\alpha )^{N-1} 
    \notag \\
    &\phantom= + \{ Q_{12} Q_3 + Q_{23} Q_1 + Q_{31} Q_2 \} s^2 ( s-\alpha )^{N-2} 
    \notag \\
    &\phantom= + \{ Q_1 Q_2 Q_3 \} s^3 ( s-\alpha )^{N-3} 
    \Big] ,
    \label{eq:d3G~m0}
\end{align}
and so forth, where we introduced the following notations:
\begin{align}
    Q_{w_1w_2} 
    &\equiv \sum_i \xi_i^{(w_1)} \xi_i^{(w_2)} ,
    \label{eq:Q12} \\
    Q_{w_1w_2w_3} 
    &\equiv \sum_i \xi_i^{(w_1)} \xi_i^{(w_2)} \xi_i^{(w_3)} ,
    \\
    \{ Q_{w_1} Q_{w_2} \}
    &\equiv \sum_{i\ne j} \xi_i^{(w_1)} \xi_j^{(w_2)}
    = Q_{w_1} Q_{w_2} - Q_{w_1w_2} ,
    \\
    \{ Q_{w_1w_2} Q_{w_3} \}
    &\equiv \sum_{i\ne j} \xi_i^{(w_1)} \xi_i^{(w_2)} \xi_j^{(w_3)}
    = Q_{w_1w_2} Q_{w_3} - Q_{w_1w_2w_3} ,
    \\
    \{ Q_{w_1} Q_{w_2} Q_{w_3} \}
    &\equiv \sum_{i\ne j,j\ne k,k\ne i} \xi_i^{(w_1)} \xi_j^{(w_2)} \xi_k^{(w_3)}
    \notag \\
    &= Q_{w_1} Q_{w_2} Q_{w_3} - ( Q_{w_1w_2} Q_{w_3} + Q_{w_2w_3} Q_{w_1} + Q_{w_3w_1} Q_{w_2} ) - Q_{w_1w_2w_3} ,
    \label{eq:Q1Q2Q3}
\end{align}
etc. 

To deal with the second term in Eq.~\eqref{eq:d2G~m0}, we extend the generating function~\eqref{eq:G~m} so that it contains another source term as
\begin{align}
    \tilde G( s;\theta_1,\theta_2,\theta_{12} ) = \sumint{\tilde P} s^n e^{q_1\theta_1+q_2\theta_2+q_{12}\theta_{12}} ,
    \label{eq:G~1212}
\end{align}
where $q_{12}=\sum_{i\in\rm (observed)} \xi_i^{(1)} \xi_i^{(2)}$ is regarded as a new physical quantity and $\theta_{12}$ is its source term. The other notations for $q$, such as $q_{123}$ and $\{q_1q_2\}$, are also introduced similarly to Eqs.~\eqref{eq:Q12}--\eqref{eq:Q1Q2Q3} for observed particles. One then finds that 
\begin{align}
    \frac{\partial \tilde G}{\partial\theta_{12} }\Big|_{\theta=0}
    &= \sumint{P} r^N Q_{12} s ( s-\alpha )^{N-1} ,
    \label{eq:d1G~m12}
\end{align}
with the same procedure as Eq.~\eqref{eq:d1G~m0}, and hence
\begin{align}
    \sumint{P} r^N \{ Q_1 Q_2 \} s^2 ( s-\alpha )^{N-2} 
    = \frac{\partial^2 \tilde G}{\partial\theta_1\partial\theta_2}\Big|_{\theta=0} - \frac{\partial \tilde G}{\partial\theta_{12}}\Big|_{\theta=0} .
    \label{eq:Q1Q2=dG~}
\end{align}
Dividing both sides of Eq.~\eqref{eq:Q1Q2=dG~} by $s^2$ and taking the double integral, one obtains 
\begin{align}
    \Big\langle \frac{ \{Q_1Q_2\} }{ N(N-1) } \Big\rangle
    &= \int_\alpha^1 ds' \int_\alpha^{s'} ds \sumint{P} r^N \{Q_1Q_2\} (s-\alpha)^{N-2}
    \notag \\
    &= \int_\alpha^1 ds' \int_\alpha^{s'} ds \frac1{s^2} \Big( \frac{\partial^2 \tilde G}{\partial\theta_1\partial\theta_2}\Big|_{\theta=0} - \frac{\partial \tilde G}{\partial\theta_{12}}\Big|_{\theta=0} \Big) .
\end{align}
Substituting Eq.~\eqref{eq:G~1212} into this result, one obtains
\begin{align}
    \Big\langle \frac{ \{Q_1Q_2\} }{ N(N-1) } \Big\rangle
    &= \sumint{\tilde P,n\ne0,1} ( q_1 q_2 - q_{12} ) \int_\alpha^1 ds' \int_\alpha^{s'} ds s^{n-2} 
    \notag \\
    &= \Bigllangle \frac{\{ q_1 q_2 \}}{n(n-1)} \kappa_{2;n}\Bigrrangle_{n\ne0,1} ,
    \label{eq:correctionH2} \\
    \kappa_{2;n} 
    &= \int_\alpha^1 ds' \int_\alpha^{s'} ds s^{n-2} 
    = 1- n \alpha^{n-1} + (n-1)\alpha^n .
    \label{eq:kappa2}
\end{align}
This result gives the formula for the efficiency correction of second-order correlation functions. 
We notice that $Q_1$ and $Q_2$ can be the same physical quantity. The right-hand side of Eq.~\eqref{eq:correctionH2} is not $\llangle \{ q_1 q_2 \}/n(n-1)\rrangle_{n\ne0,1}$, but contains additional terms arising from the second and third terms of Eq.~\eqref{eq:kappa2}. From the same procedure as Eq.~\eqref{eq:qnalpha}, one can show that these terms correspond to the contributions of the $n=0,1$ events.

A similar manipulation leads to the result for the third-order correlation function 
\begin{align}
    \Big\langle \frac{ \{Q_1Q_2Q_3\} }{ N(N-1)(N-2) } \Big\rangle
    &= \Bigllangle \frac{\{ q_1 q_2 q_3 \}}{n(n-1)(n-2)} \kappa_{3;n} \Bigrrangle_{n\ne0,1,2},
    \label{eq:correctionH3}
    \\
    \kappa_{3;n} 
    &= \int_\alpha^1 ds_1 \int_\alpha^{s_1} ds_2 \int_\alpha^{s_2} ds_3 s_3^{n-3} 
    \notag \\
    &= 1- \frac{n(n-1)}2 \alpha^{n-2} + n(n-2) \alpha^{n-1} - \frac{(n-1)(n-2)}2 \alpha^n ,
    \label{eq:kappa3}
\end{align}
and yet higher orders, whereas the manipulation becomes lengthy.

\subsection{Multiple efficiencies}
\label{sec:higher_mult}

Next, we consider the reconstruction of higher-order correlations for the case where the efficiencies are not uniform. 

To address this problem, as we did in Sec.~\ref{sec:1st_mult} we assume that our detector is divided into multiple efficiency bins having different efficiencies $r_1,r_2,\cdots,r_A$, where $A$ is the number of bins. We then denote the number of particles arriving at $a$th efficiency bin as $N_a$, and their physical quantities as $\xi_i^{(w,a)}$ ($i=1,2,\cdots,N_A)$, where $w$ is the label representing different physical quantities as in the previous subsection and the subscript $i$ represents different particles in the efficiency bin $a$. We denote their probability distribution as $P(\{N_a;\vec\xi^{(w,a)}\})$. 

After the measurement of this system with efficiency loss, we may find $n_a$ particles in the $a$th efficiency bin and the sum of the physical quantities $q_{w,a}=\sum_{i\in\rm observed} \xi_i^{(w,a)}$. We denote this probability distribution function as $\tilde P(\{n_a;q_{w,a}\})$.

To deal with multiple efficiency bins, in this section we also modify Eq.~\eqref{eq:Q12}--\eqref{eq:Q1Q2Q3} as
\begin{align}
    Q_w 
    &= \sum_{a=1}^A Q_{w,a} 
    \equiv \sum_{a=1}^A \sum_{i=1}^{N_a} \xi_i^{(w,a)} ,
    \\
    Q_{w_1w_2} 
    &= \sum_{a=1}^A \sum_{i=1}^{N_a} \xi_i^{(w_1,a)}\xi_i^{(w_2,a)} ,
    \\
    \{Q_{w_1}Q_{w_2}\} 
    &= Q_{w_1} Q_{w_2} - Q_{w_1w_2}
    = \sum_{a,b=1}^A \sum_{i=1}^{N_a} \sum_{j=1}^{N_b}  \xi_i^{(w_1,a)}\xi_j^{(w_2,b)} (1-\delta_{a,b}\delta_{i,j}) ,
\end{align}
etc. The notations for observed quantities, such as $q_{w_1w_2}=\sum_a \sum_{i\in{\rm(observed)}}\xi_i^{(w_1,a)}\xi_i^{(w_2,a)}$ are also extended accordingly.

We then repeat the same procedure as before, which leads us to the following representation of the generating function of $\tilde P(\{n_a;q_{w,a}\})$,
\begin{align}
    \tilde G(\{s_a;\theta_{w,a}\})
    &= \sumint{\tilde P} \prod_{a=1}^A s_a^{N_a} e^{\sum_w \theta_{w,a} q_{w,a}} e^{\sum_{w_1,w_2} \theta_{w_1w_2,a} q_{w_1w_2,a}} \cdots
    \label{eq:G~wa} \\
    &= \sumint{P} \prod_{a=1}^A \prod_{i=1}^{N_a} \big( 1-r_a + r_a s_a e^{\sum_w \theta_{w,a} \xi_i^{(w,a)}} e^{\sum_{w_1,w_2} \theta_{w_1w_2,a} \xi_i^{(w_1,a)}\xi_i^{(w_2,a)}} \cdots \big) ,
    \label{eq:G~wa2}
\end{align}
with $q_{12,a}= \sum_{i\in{\rm(observed)}}\xi_i^{(1,a)}\xi_i^{(2,a)}$, where $\theta_{w,a}$ and $\theta_{w_1w_2,a}$ are the external parameters associated to $q_{w,a}$ and $q_{w_1w_2,a}$. The contribution of higher-order terms $q_{123,a}= \sum_{i\in{\rm(observed)}}\xi_i^{(1,a)}\xi_i^{(2,a)}\xi_i^{(3,a)}$, which are necessary to manipulate the higher-order correlations are abbreviated in Eq.~\eqref{eq:G~wa2}.

The second derivative of Eq.~\eqref{eq:G~wa2} is calculated to be
\begin{align}
    \sum_{a,b=1}^A \frac{\partial_{1,a}}{r_a s_a} \frac{\partial_{2,b}}{r_b s_b} \tilde G \Big|_{\theta=0}
    &= \sumint{P} \bigg[ \sum_{a=1}^A \sum_{i=1}^{N_a} \frac{\xi_i^{(1,a)}\xi_i^{(2,a)}}{r_a s_a} \sigma^{\sum_a N_a-1}
    \notag \\
    &\phantom=
    + \sum_{a,b=1}^A \sum_{i=1}^{N_a} \sum_{j=1}^{N_b}  \xi_i^{(1,a)}\xi_j^{(2,b)} (1-\delta_{a,b}\delta_{i,j}) \sigma^{\sum_a N_a-2}\bigg] ,
    \label{eq:d2G~}
\end{align}
with $\partial_{w,a}=\partial/\partial\theta_{w,a}$, where we replaced $s_a$ by $\sigma$ according to Eqs.~\eqref{eq:s-sigma} and~\eqref{eq:sigma-s} after taking the derivatives.
The term $(1-\delta_{a,b}\delta_{i,j})$ in the last term of Eq.~\eqref{eq:d2G~} means that the contribution from the identical particle ($a=b$ and $i=j$) is removed from the summation.
From Eq.~\eqref{eq:d2G~}, one then obtains
\begin{align}
    \sumint{P} \{Q_1Q_2\}\sigma^{N-2}
    &=
    \sum_{a,b=1}^A \frac{\partial_{1,a}}{r_a s_a} \frac{\partial_{2,b}}{r_b s_b} \tilde G \Big|_{\theta=0}
    - \sum_{a=1}^A \frac{\partial_{12,a}}{r_a^2 s_a^2} \tilde G \Big|_{\theta=0} ,
    \label{eq:Q1Q2=dG}
\end{align}
with $\partial_{w_1w_2,a}=\partial/\partial\theta_{w_1w_2,a}$.

By taking the double integral of both sides of Eq.~\eqref{eq:Q1Q2=dG} and repeating the same procedure as before using Eq.~\eqref{eq:G~wa}, we find the reconstruction formula for the second-order correlation
\begin{align}
    \Big\langle \frac{\{Q_1Q_2\}}{N(N-1)} \Big\rangle 
    &=
    \int_0^1 ds' \int_0^{s'} ds \bigg[ \sum_{a,b=1}^A \frac{\partial_{1,a}}{r_a s_a} \frac{\partial_{2,b}}{r_b s_b} \tilde G \Big|_{\theta=0}
    - \sum_{a=1}^A \frac{\partial_{12,a}}{r_a^2 s_a^2} \tilde G \Big|_{\theta=0} \bigg]
    \notag \\
    &= \Bigllangle \sum_{a,b} \{q_{1,a}q_{2,b}\} K_{2;a,b} \Bigrrangle_{n\ne0,1} ,
    \label{eq:correctionC2K}
\end{align}
with
\begin{align}
    K_{2;a,b}
    &= \frac1{r_a r_b} \int_0^1 d\sigma' \int_0^{\sigma'} d\sigma \frac1{(\sigma/r_a+\alpha_a)(\sigma/r_b+\alpha_b)} \prod_{c=1}^A \Big( \frac\sigma{r_c} + \alpha_c \Big)^{n_c} ,
    \label{eq:K2}\\
    \{q_{1,a}q_{2,b}\} 
    &= \sum_{a,b=1}^A \sum_{i,j\in\rm (observed)}  \xi_i^{(1,a)}\xi_j^{(2,b)} (1-\delta_{a,b}\delta_{i,j}) 
    = q_{1,a} q_{2,b} - \delta_{ab} q_{12,a} ,
\end{align}
and $\alpha_a=(r_a-1)/r_a$.
For the case where the efficiencies are different for every particle, these results are rewritten as 
\begin{align}
    \Big\langle \frac{\{Q_1Q_2\}}{N(N-1)} \Big\rangle 
    &= \Bigllangle \sum_{i\ne j} q_{1,i}q_{2,j} k_{2;i,j} \Bigrrangle_{n\ne0,1} ,
    \label{eq:correctionC2} \\
    k_{2;i,j}
    &= \frac1{r_i r_j} \int_0^1 d\sigma' \int_0^{\sigma'} d\sigma \prod_{l\ne i,l\ne j} \frac{\sigma + r_l \alpha_l}{r_l} 
    \notag \\
    &= \Big( \prod_i r_i \Big)^{-1} \int_0^1 d\sigma' \int_0^{\sigma'} d\sigma \prod_{l\ne i,l\ne j} ( \sigma + r_l - 1 ) ,
    \label{eq:k2}
\end{align}

Similar manipulations allow one to extend the formula to higher-order correlations as
\begin{align}
    \Big\langle \frac{\{Q_1Q_2Q_3\}}{N(N-1)(N-2)} \Big\rangle 
    &= \Bigllangle \sum_{i\ne j,j\ne k,k\ne i} q_{1,i}q_{2,j} q_{3,k} k_{2;i,j,k} \Bigrrangle_{n\ne0,1,2} ,
    \label{eq:correctionC3}
    \\
    k_{3;i,j,k}
    &= \Big( \prod_i r_i \Big)^{-1} \int_0^1 d\sigma'' \int_0^{\sigma''} d\sigma' \int_0^{\sigma'} d\sigma \prod_{l\ne \{i,j,k\}} ( \sigma + r_l - 1 ) ,
    \label{eq:k3}
\end{align}
and so forth.

\section{Discussions}

In this study, we derived the analytic formulas for the efficiency correction of particle-averaged quantities and their higher-order correlations, Eqs.~\eqref{eq:correction3}, \eqref{eq:correctionC2}, and~\eqref{eq:correctionC3}. These results do not agree with the conventional formulas~\eqref{eq:conventional}. 

There are some remaining issues to be resolved in the future study. Although we have obtained the formulas for the efficiency correction, they take rather complicated forms involving integrals, and their intuitive interpretation is not straightforward. For instance, the relation of these results to the conventional formulas~\eqref{eq:conventional} is unclear. Rewriting them into more transparent forms is an important task that the authors have not accomplished. In addition, developing efficient numerical procedures for their evaluations will also be important for practical purposes. The use of the relations between elementary symmetric polynomials, which are partly discussed in Sec.~\ref{sec:1st_comment}, will be helpful in this context.

In Sec.~\ref{sec:higher}, we studied the efficiency correction of the higher-order correlations~\eqref{eq:corr2} and~\eqref{eq:corr3} that do not contain the self-correlations. However, the correlations containing the self-correlation, such as $( \sum_i\xi_i/N)^2$, would be more useful for some purposes. To extend our results for their efficiency corrections, we need additional procedures, which will be reported in a future study. In some experimental studies, a second-order correlation is defined through two-particle correlations between completely different acceptance regions. Extension of the reconstruction formulas to this case is another issue that has not been discussed in this paper, which, however, is rather straightforward.

\section*{Acknowledgment}

The authors thank Jianyong Jia, Xiaofeng Luo and You Zhou for valuable discussions. 
This work was supported in part by JSPS KAKENHI (Nos.~JP19H05598, JP22K03619, JP23H04507, JP23K13113, JP24K07049), ISHIZUE 2025 of Kyoto University, and the Center for Gravitational Physics and Quantum Information (CGPQI) at Yukawa Institute for Theoretical Physics.

\appendix

\section{Reconstruction of $\langle 1/N\rangle$}
\label{app:1/N}

In this appendix, we address a simplified problem as an exercise of the main text. We consider a positive-integer stochastic variable $N>0$, whose distribution is given by the probability distribution function $P(N)$.
The variable $N$ may be interpreted as the event-by-event particle number. We then consider a problem of reconstructing the expectation value of the inverse $N$
\begin{align}
    \Big\langle \frac1N \Big\rangle = \sum_{N=1}^\infty P(N) \frac1N .
    \label{eq:1/N}
\end{align}

We assume that every particle is observed with imperfect probability, i.e., efficiency, $r$, which is independent for individual particles. In this situation, the distribution function $\tilde P(n)$ of observed particle number $n$ is obviously different from $P(N)$. Now we try to obtain the correct value of Eq.~\eqref{eq:1/N} from the imperfect experimental result on $\tilde P(n)$. Notice that $P(0)=0$ so that $1/N$ is always meaningful.

To address this problem, we use the fact that $\tilde P(n)$ is related to $P(N)$ as~\cite{Kitazawa:2011wh,Kitazawa:2012at,Asakawa:2015ybt}
\begin{align}
    \tilde P(n) = \sum_N P(N) \binom Nn r^n (1-r)^{N-n} .
    \label{eq:tildeP(n)}
\end{align}
Then, the factorial generating function of $\tilde P(n)$ is given by
\begin{align}
    \tilde G(s) 
    &= \sum_n \tilde P(n) s^n 
    = \sum_N P(N) \sum_n \binom Nn (sr)^n (1-r)^{N-n} 
    \notag \\
    &= \sum_N P(N) ( 1-r+rs)^N =G(1-r+rs),
    \label{eq:tildeG(s)}
\end{align}
where we used Eq.~\eqref{eq:tildeP(n)} at the second equality and 
\begin{align}
    G(s) = \sum_N P(N) s^N ,
\end{align}
is the factorial generating function of $P(N)$. We also notice
\begin{align}
    G(0) = \tilde G(\alpha) = 0,
    \label{eq:G=G=0}
\end{align}
with $\alpha=(r-1)/r$.

From Eq.~\eqref{eq:tildeG(s)}, one finds the following relation 
\begin{align}
    \int_\alpha^1 ds \frac{\tilde G(s)}{s-\alpha} 
    =& \, \sum_N P(N) \int_\alpha^1 ds r(1-r+rs)^{N-1} = \Big\langle \frac1N \Big\rangle ,
    \label{eq:int1N}
\end{align}
while the same integral~\eqref{eq:int1N} is calculated to be
\begin{align}
    \int_\alpha^1 ds \frac{\tilde G(s)}{s-\alpha} 
    = \sum_n \tilde P(n) \int_\alpha^1 ds \frac{s^n}{s-\alpha} 
    \equiv \Bigllangle \int_\alpha^1 ds \frac{s^n}{s-\alpha} \Bigrrangle,
    \label{eq:int1N2}
\end{align}
using the first equality of Eq.~\eqref{eq:tildeG(s)}.
Equations~\eqref{eq:int1N} and~\eqref{eq:int1N2} suggest 
$\langle 1/N \rangle = \llangle \int_\alpha^1 ds s^n/(s-\alpha)\rrangle$, which is the formula to represent $\langle 1/N \rangle$ 
in terms of $\tilde P(n)$. 
However, the integral on the right-hand side is divergent, and this equation does not make sense. To remove this divergence, we regularize the integral in Eq.~\eqref{eq:int1N2} using Eq.~\eqref{eq:G=G=0} as 
\begin{align}
    \int_\alpha^1 ds \frac{\tilde G(s)}{s-\alpha} 
    =\int_\alpha^1 ds \frac{\tilde G(s)-\tilde G(\alpha)}{s-\alpha} 
    = \sum_n \tilde P(n) \int_\alpha^1 ds \frac{s^n-\alpha^n}{s-\alpha} .
\end{align}
This leads to
\begin{align}
    \Big\langle \frac1N \Big\rangle 
    = \llangle K_{n} \rrangle,
    \label{eq:1/N=K}
\end{align}
with
\begin{align}
    K_n = \int_\alpha^1 ds \frac{s^n-\alpha^n}{s-\alpha} .
    \label{eq:K_n}
\end{align}
Equation~\eqref{eq:1/N=K} is the answer to our problem, i.e. $\langle 1/N \rangle$ is represented in terms of quantities constructed from $\tilde P(n)$. Because $K_0=0$, the summation in Eq.~\eqref{eq:1/N=K} is not taken for $n=0$.

Since the integrand in Eq.~\eqref{eq:K_n} is given by a polynomial of $s$, the integral can be calculated analytically. However, for practical purposes it may be more robust and easier to calculate it numerically.

\subsection{Check of Eq.~\eqref{eq:1/N=K}}

To check the validity of Eq.~\eqref{eq:1/N=K}, we consider a simple distribution function
\begin{align}
    P(N) = \delta_{N,N_0} ,
    \label{eq:P(N)0}
\end{align}
that the value of $N$ is fixed to $N_0$. Equaiton~\eqref{eq:P(N)0} of course gives $\langle1/N\rangle=1/N_0$.

From Eq.~\eqref{eq:tildeP(n)}, the ``observed'' distribution $\tilde P(n)$ corresponding to Eq.~\eqref{eq:P(N)0} is given by 
\begin{align}
    \tilde P(n) = \binom {N_0}n (1-r)^{N_0-n} r^n
    = \binom {N_0}n r^{N_0} (-\alpha)^{N_0-n}.
    \label{eq:tildeP(n)0}
\end{align}
For Eq.~\eqref{eq:tildeP(n)0}, the right-hand side of Eq.~\eqref{eq:1/N=K} is calculated to be
\begin{align}
    \llangle K_n \rrangle 
    =&\, \sum_n \tilde P(n) K_n
    \notag \\
    =&\, 
    r^{N_0} \int_\alpha^1 ds \sum_n \frac{s^n-\alpha^n}{s-\alpha} \binom {N_0}n  (-\alpha)^{N_0-n}
    \notag \\
    =&\, 
    (-r\alpha)^{N_0} \int_\alpha^1 ds \frac1{s-\alpha} \sum_n \binom {N_0}n  \Big( \Big(-\frac s\alpha\Big)^n-(-1)^n \Big)
    \notag \\
    =&\, 
    r^{N_0} \int_\alpha^1 ds (s-\alpha)^{N_0-1} \notag \\
    =&\, \frac1{N_0},
\end{align}
which gives the correct answer. 

One can easily check that the result is valid even for general $P(N)$ by extending the above analysis.

\section{Efficiency correction in a simple model}
\label{app:binom}

In this appendix, we consider a simple model to check the validity of the reconstruction formulas, Eqs.~\eqref{eq:correction2} and~\eqref{eq:correctionC2K}.

We consider a physical quantity taking only two values $\xi_i=\pm1$. It is also assumed that the particle numbers having $\xi_i=\pm1$ are fixed to $N_\pm$, respectively, in all events. We thus have 
\begin{align}
    \Big\langle \frac QN \Big\rangle &= \frac1N \sum_i \xi_i = \frac{N_+-N_-}{N_++N_-} ,
    \label{eq:QNpm} \\       \Big\langle \frac{\{QQ\}}{N(N-1)} \Big\rangle &= \frac{(N_+-N_-)^2-(N_++N_-)}{(N_++N_-)(N_++N_--1)} ,
    \label{eq:QNpm2}
\end{align}
as a result of the perfect measurement with the total particle number $N=N_++N_-$. Although this is an artificial model for a demonstration, one may regard the physical quantity as the net-baryon number, and $N_+$ baryons and $N_-$ anti-baryons are produced in all events.

We then assume that the particles with $\xi_i=\pm1$ are observed with the efficiencies $r_\pm$, respectively.
In this case, the probability of observing $n_\pm$ particles, respectively, is given by
\begin{align}
    \tilde p(n_+,n_-) = \binom {N_+}{n_+} r_+^{n_+} (1-r_+)^{N_+-n_+} \binom {N_-}{n_-} r_-^{n_-} (1-r_-)^{N_--n_-} .
\end{align}

\begin{figure}
    \centering
    \includegraphics[width=0.5\linewidth]{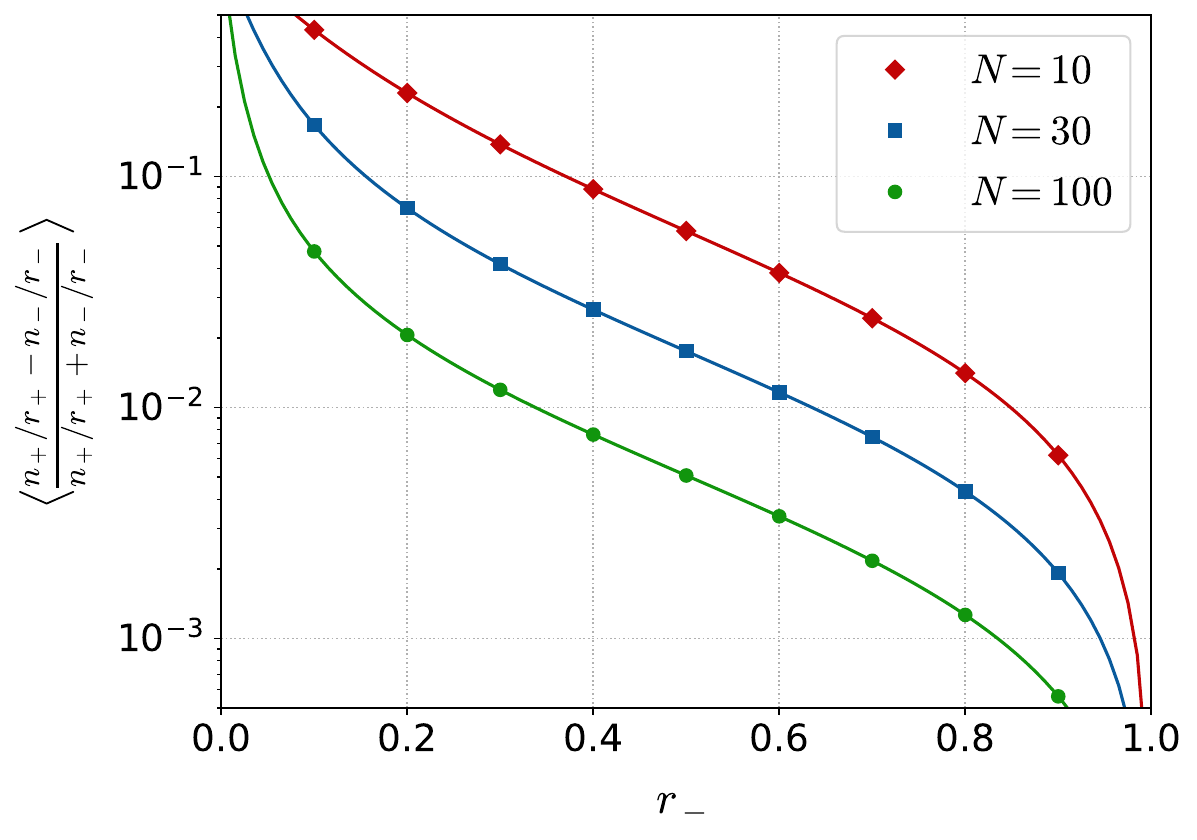}
    \caption{Reconstructed value with the conventional formula~\eqref{eq:<<pm>>} for $N_+=N_-=N/2$ and $r_+=1$ as a function of $r_-$ for three values of $N$.}
    \label{fig:pm}
\end{figure}

Equation~\eqref{eq:correction2} in this system reads 
$\langle Q/N \rangle = \llangle q_+ K_{1;+} + q_- K_{1;-} \rrangle_{n\ne0}$ with 
\begin{align}
    q_\pm &= \pm n_\pm ,
    \\
    K_{1;\pm} &= \int_0^1 d\sigma \frac1{r_\pm} \Big( \frac\sigma{r_\pm} + \alpha_\pm \Big)^{n_\pm-1}
    \Big( \frac\sigma{r_\mp} + \alpha_\mp \Big)^{n_\mp} .
\end{align}
Hence, it is calculated to be
\begin{align}
    &\llangle q_+ K_{1;+} + q_- K_{1;-}\rrangle_{n\ne0} 
    \notag \\
    &= \sum_{n_+,n_-}\int_0^1 d\sigma \bigg[ 
    \frac{n_+}{r_+} \Big( \frac\sigma{r_+} + \alpha_+ \Big)^{n_+-1}
    \Big( \frac\sigma{r_-} + \alpha_- \Big)^{n_-} 
    + (\pm\to\mp) \bigg] \tilde p(n_+,n_-)
    \notag \\
    &= \int_0^1 d\sigma \bigg[
    \sum_{n_+} \frac{n_+}{r_+} \Big( \frac\sigma{r_+} + \alpha_+ \Big)^{n_+-1} \binom {N_+}{n_+} r_+^{n_+} (1-r_+)^{N_+-n_+} 
    \notag \\
    &\phantom{=\int_0^1 d\sigma}
    \times \sum_{n_-}
    \Big( \frac\sigma{r_-} + \alpha_- \Big)^{n_-} \binom {N_-}{n_-} r_-^{n_-} (1-r_-)^{N_--n_-}     
    + (\pm\to\mp) \bigg] 
    \notag \\
    &= \int_0^1 d\sigma \Big[ N_+ \sigma^{N_+-1} \sigma^{N_-} - N_- \sigma^{N_--1} \sigma^{N_+} \Big]
    \notag \\
    &= \frac{ N_+ - N_- }{ N_+ + N_- } ,
\end{align}
which reproduces Eq.~\eqref{eq:QNpm} as it should be. 

For the second-order correlation, we have 
\begin{align}
    \{ q_\pm q_\pm \} = n_\pm(n_\pm-1), 
    &&
    \{ q_+ q_- \} = q_+q_- = -n_+n_-,
\end{align}
\begin{align}
    K_{2;\pm\pm} 
    &= \int_0^1 d\sigma' \int_0^1 d\sigma \frac1{r_\pm^2} \Big( \frac\sigma{r_\pm} + \alpha_\pm \Big)^{n_\pm-2}
    \Big( \frac\sigma{r_\mp} + \alpha_\mp \Big)^{n_\mp} ,
    \\
    K_{2;+-} 
    &= \int_0^1 d\sigma' \int_0^1 d\sigma \frac1{r_+r_-} \Big( \frac\sigma{r_+} + \alpha_+ \Big)^{n_+-1}
    \Big( \frac\sigma{r_-} + \alpha_- \Big)^{n_--1},
\end{align}
and Eq.~\eqref{eq:correctionC2K} is rewritten as
\begin{align}
    &
    \Bigllangle
    \{ q_+q_+\} K_{2;++} + 2q_+q_- K_{2;+-} + \{q_-q_-\} K_{2;--} \Bigrrangle
    \notag \\
    &=
    \sum_{n_+,n_-} \tilde p(n_+,n_-) \Big\{\tilde
    n_+(n_+-1) K_{2;++} - 2n_+n_- K_{2;+-} + n_-(n_--1) K_{2;--} \Big\} .
    \label{eq:2nd+-}
\end{align}
Repeating a similar manipulation as above, one can show that Eq.~\eqref{eq:2nd+-} reproduces the correct answer, Eq.~\eqref{eq:QNpm2}.
Similar manipulations are also extended to yet higher-order correlations.
It is not difficult to show the validity of Eqs.~\eqref{eq:correction2} and~\eqref{eq:correction3} in more complicated situations by extending the above argument.

Finally, let us see the reconstruction with the conventional formula~\eqref{eq:conventional} in this model for Eq.~\eqref{eq:QNpm}. 
According to Eq.~\eqref{eq:conventional}, the reconstructed value in this model reads
\begin{align}
    \Bigllangle \frac{n_+/r_+ - n_-/r_- }{n_+/r_+ + n_-/r_- } \Bigrrangle.
    \label{eq:<<pm>>}
\end{align}
In order to see if this formula reproduces Eq.~\eqref{eq:QNpm}, now we consider a simple case with $N_+=N_-=N/2$, which gives $\langle Q/N \rangle = \langle (N_+-N_-)/(N_++N_-) \rangle =0$. In Fig.~\ref{fig:pm}, we show the value of Eq.~\eqref{eq:<<pm>>} for $r_+=1$ as a function of $r_-$ for several values of $N$. The figure shows that the value of Eq.~\eqref{eq:<<pm>>} is nonzero except for $r_-=1$. This result shows that Eq.~\eqref{eq:<<pm>>} does not reconstruct the correct result $\langle Q/N \rangle=0$ in this case, while the reconstructed value approaches the correct result for $N\to\infty$ or $r_-\to1$.

\let\doi\relax


\bibliographystyle{unsrt}
\bibliography{references}

\end{document}